\documentclass[apjl]{emulateapj}
\usepackage{apjfonts}
\begin{document}
\shorttitle{Constraining RIAF Models of Sgr~A* with VLBI}
\shortauthors{Fish et al.}
\title{Using Millimeter VLBI to Constrain RIAF Models of Sagittarius~A*}
\author{Vincent L.\ Fish\altaffilmark{1},
        Avery E.\ Broderick\altaffilmark{2},
        Sheperd S.\ Doeleman\altaffilmark{1}, \&
        Abraham Loeb\altaffilmark{3}
}
\altaffiltext{1}{Massachusetts Institute of Technology, Haystack
  Observatory, Route 40, Westford, MA 01886; vfish@haystack.mit.edu,
  dole@haystack.mit.edu}
\altaffiltext{2}{Canadian Institute for Theoretical Astrophysics,
  University of Toronto, 60 St.\ George St., Toronto, ON, M5S 3H8
  Canada; aeb@cita.utoronto.ca}
\altaffiltext{3}{Institute for Theory and Computation, Harvard
  University, Center for Astrophysics, 60 Garden St., Cambridge, MA
  02138; aloeb@cfa.harvard.edu}

\begin{abstract}
The recent detection of Sagittarius~A* at $\lambda = 1.3$~mm on a
baseline from Hawaii to Arizona demonstrates that millimeter
wavelength very long baseline interferometry (VLBI) can now spatially
resolve emission from the innermost accretion flow of the Galactic
center region.  Here, we investigate the ability of future millimeter
VLBI arrays to constrain the spin and inclination of the putative
black hole and the orientation of the accretion disk major axis within
the context of radiatively inefficient accretion flow (RIAF) models.
We examine the range of baseline visibility and closure amplitudes
predicted by RIAF models to identify critical telescopes for
determining the spin, inclination, and disk orientation of the Sgr~A*
black hole and accretion disk system.  We find that baseline lengths
near $3$~G$\lambda$ have the greatest power to distinguish amongst
RIAF model parameters, and that it will be important to include new
telescopes that will form north-south baselines with a range of
lengths.  If a RIAF model describes the emission from Sgr~A*, it is
likely that the orientation of the accretion disk can be determined
with the addition of a Chilean telescope to the array.  Some likely
disk orientations predict detectable fluxes on baselines between the
continental United States and even a single 10--12~m dish in Chile.
The extra information provided from closure amplitudes by a
four-antenna array enhances the ability of VLBI to discriminate
amongst model parameters.
\end{abstract}

\section{Introduction}
\label{introduction}

The Galactic center radio source Sagittarius~A* is believed to be
associated with a black hole with a mass of approximately $4 \times
10^6$~M$_\sun$ at a distance of about 8~kpc \citep{schodel03,ghez08}.
The event horizon of Sgr~A* has the largest apparent angular size of
all known black holes as viewed from Earth.  High angular resolution
is critical to understanding Sgr~A* because the size scales are so
small: for instance, the apparent lensed horizon size is $55~\mu$as.
The quest for angular resolution has driven observers toward very long
baseline interferometry (VLBI) at millimeter wavelengths.  Most
recently, \citet{doeleman08} detected Sgr~A* at 230~GHz ($\lambda =
1.3$~mm) on baselines between the James Clerk Maxwell Telescope (JCMT)
in Hawaii and the Submillimeter Telescope Observatory (SMTO) in
Arizona, as well as between the SMTO and a Combined Array for Research
in Millimeter-wave Astronomy (CARMA) telescope in California.  The
former is the longest baseline ($3.5$~G$\lambda$, fringe spacing $\sim
60~\mu$as) on which Sgr~A* has ever been detected.

In the absence of a true ab initio simulation of Sgr~A*'s accretion
flow, a variety of physically simplified models have been proposed.
Constraints from the observed spectrum, polarization, and variability
have not proved sufficient to conclusively establish the nature of the
emission region.  This is evidenced by the continuing vigorous debate
over the morphology of this region, e.g., primarily from the innermost
regions of an inefficient accretion flow \citep{yuan03}, the
acceleration region of a nascent jet \citep{falcke00}, or something
qualitatively different \citep{igumenshchev03}.  However, millimeter
VLBI's ability to resolve the emitting region directly promises to
discriminate amongst these.  To do so will require a detailed
understanding of the role that key physical parameters (such as black
hole spin, disk/jet orientation, etc.) play in shaping the millimeter
images of Sgr~A*.  Conversely, if the context of the emission can be
conclusively established, millimeter VLBI has great potential to
extract these parameters.

In a related work, \citet{broderick08} examine the ability of the
\citet{doeleman08} VLBI detections to discriminate amongst various
radiatively inefficient accretion flow (RIAF) models parameterized by
black hole spin, viewing angle, and disk orientation.  A key finding
is that the detected flux on the JCMT-SMTO baseline argues against
orientations in which the assumed accretion disk is close to face on.
However, the RIAF parameters are strongly coupled, and there are not
yet enough millimeter VLBI measurements to make strong statements
about the black hole spin, for instance.  In this work, we explore the
question of which future millimeter VLBI observations are likely to
have the greatest impact for distinguishing between RIAF models and
are therefore most likely to determine the physical parameters of the
black-hole/accretion-disk system.

\section{Disk Models and Telescopes}
\label{models}

Here we consider an ensemble of RIAF models similar to that described
in \citet{yuan03} and described in detail in \citet{doeleman08b} and
\citet{broderick08}.  These are generally characterized by the
inefficient energy transfer between the electrons and the ions.  In
particular, we model the millimeter flux from Sgr~A* as synchrotron
emission due to populations of thermal (though with considerably lower
temperatures than the ions) and nonthermal electrons in a
geometrically thick, quiescent disk containing a toroidal magnetic
field.  The thermal and nonthermal electron densities and thermal
electron temperature are solved for using measured radio,
submillimeter, and near-infrared fluxes as constraints.  Model images
as viewed from the Earth are produced using fully general relativistic
radiative transfer.

Each model has a total flux density of 2.4~Jy at $\lambda = 1.3$~mm,
in line with the integrated flux density measured by
\citet{doeleman08}.  These models are parameterized by black hole spin
($a$), inclination ($\theta = 0\degr$ for a face-on disk), and
orientation of the disk major axis on the plane of the sky ($\xi$)
measured counterclockwise from east (equivalently the position angle
of the spin axis).  (Henceforth we shall use the word ``models'' to
refer to RIAF models with these three free parameters.)  Models are
produced at spins from $a = 0.0$ to 0.9 in increments of 0.1, as well
as at 0.99 and 0.998.  The ensemble includes models at $\theta =
1\degr$ and then at $\theta = 10\degr$ to $90\degr$ in $10\degr$
increments.  This particular ensemble is described in greater detail
in \citet{broderick08}.  We consider models with $ 0\degr \leq \xi <
180\degr$ with an increment of $\Delta \xi = 5\degr$ (in \S
\ref{regions}) or $30\degr$ (in \S\S \ref{baselines} and
\ref{closures}).  Visibility amplitudes are symmetric under $180\degr$
rotation for any real intensity distribution.

We consider seven potential stations for millimeter VLBI: Hawaii,
consisting of one or more of the JCMT, Submillimeter Array (SMA), and
Caltech Submillimeter Observatory (CSO) possibly phased together into
a single aperture; the SMTO on Mount Graham, Arizona; CARMA telescopes
in California, either individually or phased together; the Large
Millimeter Telescope (LMT) on Sierra Negra, Mexico; a Chilean station
consisting of either the Atacama Submillimeter Telescope Experiment
(ASTE), Atacama Pathfinder Array (APEX), or a phased array of Atacama
Large Millimeter Array (ALMA) dishes; the Institut de radioastronomie
millim\'{e}trique (IRAM) 30-m dish on Pico Veleta (PV), Spain; and the
IRAM Plateau de Bure (PdB) Interferometer, phased together as a single
aperture.  Assumed telescope capabilities and sensitivities are
detailed in \citet{doeleman08b}.

\section{Black Hole Parameter Estimation}

\subsection{Favored Regions of the $(u,v)$ Plane}
\label{regions}

For the purpose of distinguishing different models, it is most
desirable to obtain observations of Sgr~A* on baselines at points at
which the visibilities are most disparate.  We compute model
visibilities
\[
V(u,v) = \int\!\!\!\int dx\,dy\,I(x,y)\,e^{-2 \pi i (xu + yv)/\lambda},
\]
where $I(x,y)$ is the model intensity distribution with $x$ and $y$
are aligned east and north, respectively.  We characterize the
variation in model visibilities via the standard deviation
$\sigma(u,v)$ amongst all models in the ensemble.  While we initially
ignore the effects of interstellar scattering, it can be included by
multiplying the visibilities by a unit-normalized elliptical Gaussian
centered at the origin \citep[half width at half maximum of $7.0
  \times 3.8$~G$\lambda$, position angle $170\degr$ east of north,
  based on][]{bower06}.

\begin{figure}[t]
\resizebox{\hsize}{!}{\includegraphics{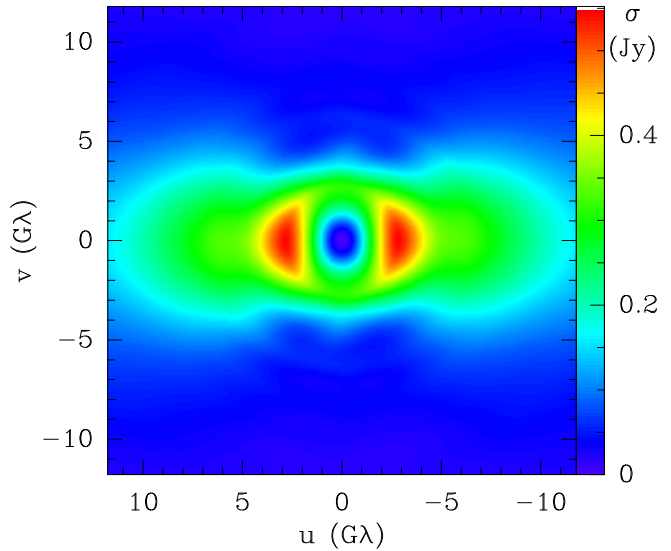}}
\caption{Map of standard deviation of predicted RIAF model
  visibilities ($\sigma(u,v)$) at $\lambda = 1.3$~mm holding $\xi =
  0\degr$ constant.  Units are in Jy.  The largest spread in predicted
  model quantities occurs when the projected baseline is aligned with
  the disk major axis.  Rotating the image on the plane of the sky
  produces a rotation of the visibility amplitudes in the $(u,v)$
  plane.
\label{fig-contrast}
}
\end{figure}

Figure~\ref{fig-contrast} shows the scatter in predicted model
visibilities, $\sigma(u,v)$.  The peak standard deviation occurs at a
baseline length of approximately $3$~G$\lambda$ at $\lambda = 1.3$~mm
($\sim 4000$~km) aligned with the accretion disk major axis.  Since
the disk orientation ($\xi$) is not known a priori, the optimal
baseline orientation is unknown.

\begin{figure}[t]
\resizebox{\hsize}{!}{\includegraphics{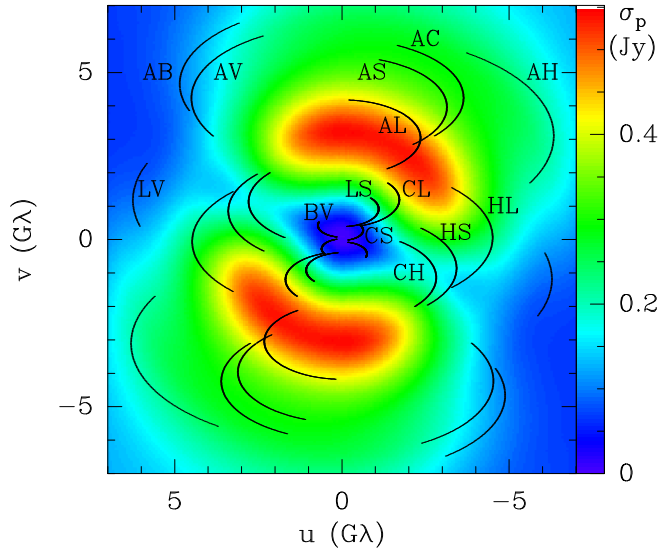}}
\caption{Plot of the standard deviation $\sigma_p(u,v)$ restricted to
  the most probable models (see \S \ref{regions}) on the inner
  part of the $(u,v)$ plane at $\lambda = 1.3$~mm.  Units are in Jy.
  Potential $(u,v)$ tracks due to Earth rotation are superposed, with
  letters indicating antennas.  A: APEX/ASTE/ALMA, B: PdB, C: CARMA,
  H: Hawaii (incl.\ JCMT), L: LMT, S: SMTO, V: PV
\label{fig-contrast2}
}
\end{figure}

Not all models are equally likely, however.  The \citet{doeleman08}
detections on the JCMT-SMTO and SMTO-CARMA baselines as well as total
flux density measurement with the CARMA interferometer place
constraints on RIAF models.  Following the approach of
\citet{broderick08}, we compute the probability $p(a,\theta,\xi)$ for
every model based on the \citet{doeleman08} detections, including a
$\sin\, \theta$ prior on the inclination.  Restricting the ensemble of
models to those with $p \geq 0.01\,p_{\mathrm{max}}$ (or approximately
one-quarter of the total set of models), we can obtain the scatter in
the predicted correlated flux densities of the most \emph{probable}
models, $\sigma_p(u,v)$ (Fig.~\ref{fig-contrast2}).  We find that the
largest scatter in models still occurs at baseline lengths near
$3$~G$\lambda$, but at orientations substantially different from that
of the JCMT-SMTO detection (labelled ``HS'' in
Fig.~\ref{fig-contrast2}), which has already placed a stringent
constraint on models in its region of $(u,v)$ space.

\subsection{Implications for Specific Baselines}
\label{baselines}

\begin{figure*}[t]
\resizebox{\hsize}{!}{\includegraphics{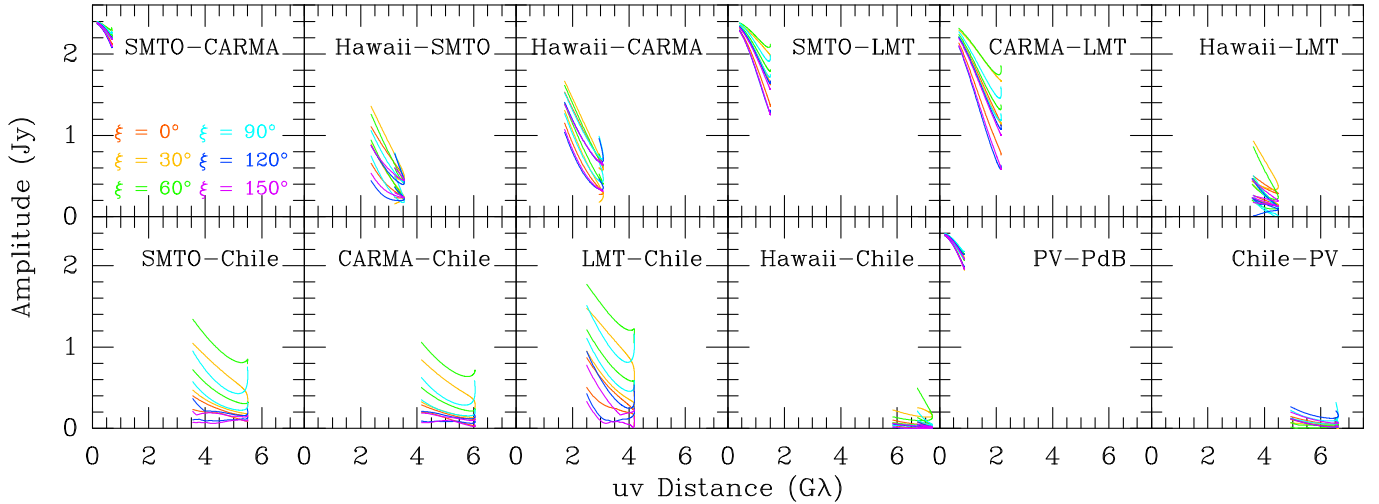}}
\caption{Predicted model fluxes on potential baselines.  Each panel
  shows the predicted track of correlated flux density with
  $\sqrt{u^2+v^2}$ (changing due to Earth rotation) for one baseline
  for all models ($\Delta\xi = 30\degr$) for which $p >
  0.01\,p_\mathrm{max}$.  Lines show the maximum and minimum predicted
  correlated flux density from the set of probable models with a given
  disk orientation, shown in color.  The visibility amplitudes on many
  baselines are highly sensitive to $\xi$.  Baselines between existing
  VLBI stations in the continental US and new stations in Chile and
  Mexico (LMT) are optimal for further constraining RIAF model
  parameter space.
\label{fig-baselines}
}
\end{figure*}

Figure~\ref{fig-baselines} shows predicted correlated flux densities
on potential millimeter VLBI baselines for the probable RIAF models.
Each pair of telescopes produces visibility measurements along a range
of projected baseline lengths due to the Earth's rotation.  These
correlated flux densities have been corrected for interstellar
scattering.

Detections have already been obtained on the Hawaii-SMTO and
SMTO-CARMA baselines, although Sgr~A* was not detected on the
Hawaii-CARMA baseline \citep{doeleman08}.  It is likely that the next
millimeter VLBI observations of Sgr~A* will occur on an array
containing these same three stations, although with the inclusion of a
phased-array processor at Hawaii to sum the signals from the JCMT,
CSO, and SMA.  The increased sensitivity should allow Sgr~A* to be
detected on the Hawaii-SMTO baseline with a higher signal-to-noise
ratio, tightening the constraints on RIAF model parameters especially
if Sgr~A* can be detected on scans when it is low to the horizon from
Arizona, where the range of predicted model flux densities is larger.
The Hawaii-CARMA baseline will be more sensitive than the Hawaii-SMTO
baseline once the planned array phasing capability is added to CARMA,
although the range in predicted model flux densities is somewhat
smaller on Hawaii-CARMA.  Because the predicted flux densities on the
Hawaii-CARMA and Hawaii-SMTO baselines are similar due to their
location in approximately the same location of the $(u,v)$ plane,
strongly constraining key black hole and accretion disk parameters
will necessitate obtaining flux density measurements on other
baselines as well.  However, measurements on the Hawaii-CARMA baseline
may be important to establish the validity of RIAF models of Sgr~A* in
general and rule out simpler, less physically-motivated models, such
as a Gaussian or a ring.

The optimal baselines for further constraining model parameter space
are SMTO-Chile, CARMA-Chile, and LMT-Chile, which are in the preferred
zone of baseline lengths and at a position angle nearly orthogonal to
that of Hawaii-SMTO.  Many of the models predict flux densities of
well over 500~mJy (some even above 1~Jy) on the SMTO-Chile and
CARMA-Chile baselines.  These flux densities should be detectable even
if the Chilean station consists of a single 10 or 12~m telescope, such
as APEX, ASTE, or a single ALMA dish.

Of the three parameters which we consider in this work, millimeter
VLBI will most easily be able to constrain the orientation of a RIAF
disk.  Disk emission comes predominantly from the Doppler-brightened
approaching side and is elongated perpendicular to the projected disk
major axis when inclined relative to the line of sight.  Because the
locations of existing and potential millimeter telescopes define
preferred directions on the sky for measuring this emission, baseline
correlated flux densities depend strongly on disk orientation.  The
scatter within the predicted correlated flux density for each baseline
is much smaller when restricted to one value of $\xi$ than when the
entire model parameter space is considered as a whole
(Fig.~\ref{fig-baselines}).

\begin{figure*}
\resizebox{\hsize}{!}{\rotatebox{-90}{\includegraphics{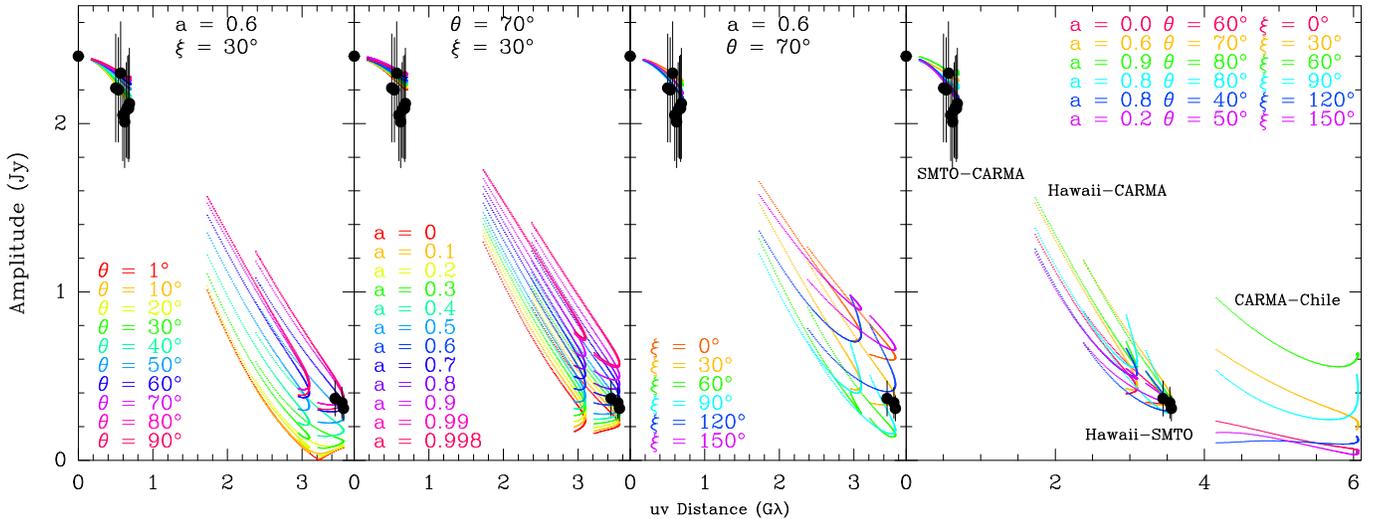}}}
\caption{\emph{Left three panels:} Correlated baseline flux density
  plots on baselines between Hawaii, SMTO, and CARMA for a likely
  model ($a = 0.6, \theta = 70\degr, \xi = 30\degr$), varying one
  parameter at a time.  Data from the second day of observations by
  \citet{doeleman08} are superposed.  Increasing the black hole spin
  and inclination predict greater flux densities on longer baselines.
  Rotation of the disk emission in the plane of the sky can cause the
  predicted flux density to increase or decrease depending on the
  baseline orientation.  \emph{Right panel:} The same for six
  different likely models showing the widely divergent predicted flux
  densities on the CARMA-Chile baseline.
\label{fig-likely}
}
\end{figure*}

Constraining the disk orientation of Sgr~A* may be a necessary first
step toward extracting the spin and inclination of the system.  For a
given orientation, increasing the black hole spin or the inclination
of the system causes the emitting region to appear smaller, raising
the correlated flux density on longer baselines
(Fig.~\ref{fig-likely}).  Indeed, if the disk orientation can be
determined, the existing \citet{doeleman08} data may already place
strong constraints on the spin and inclination \citet{broderick08}.

\subsection{Potential Complications}
\label{closures}

The flux density of 2.4~Jy at 230~GHz detected by \citet{doeleman08}
is low compared to measurements with the SMA
\citep{marrone07,marrone08} and high compared to measurements with
BIMA \citep{zhao03,bower05} and a PV-PdB VLBI measurement at 215~GHz
\citep{krichbaum98}.  Sgr~A* exhibits variability, although the
consistent fluxes measured on two consecutive days by
\citet{doeleman08} suggest that Sgr~A* was observed in its quiescent
state.  Since the variability mechanism is not understood, it remains
unclear how variability in the total flux density of Sgr~A* at
millimeter wavelengths will affect the correlated flux densities
measured on diverse VLBI baselines.  Multiple epochs of observation
will be essential to exploring the link between total flux variability
and spatial structure around Sgr~A*.  It will also be important to
obtain contemporaneous measurements of fluxes on a variety of
baselines as well as a simultaneous ``zero-spacing'' flux density,
preferably from a connected-element interferometer.

Variable and elevation-dependent antenna gains may limit the accuracy
to which telescope flux scales can be derived, introducing systematic
errors into calculations of correlated flux densities.  These
systematic errors may be difficult to characterize accurately and may
limit the ability of baseline-based quantities to distinguish between
different models.  Closure amplitudes are robust against station-based
gain variations and will provide model constraints with significantly
reduced systematic errors.  If observations are taken with at least
four telescopes, closure amplitudes can be constructed from ratios of
baseline visibilities ($A_{abcd} = |V_{ab}| |V_{cd}| |V_{ac}|^{-1}
|V_{bd}|^{-1}$, where subscripted letters identify telescopes).

\begin{figure}[t]
\resizebox{\hsize}{!}{\includegraphics{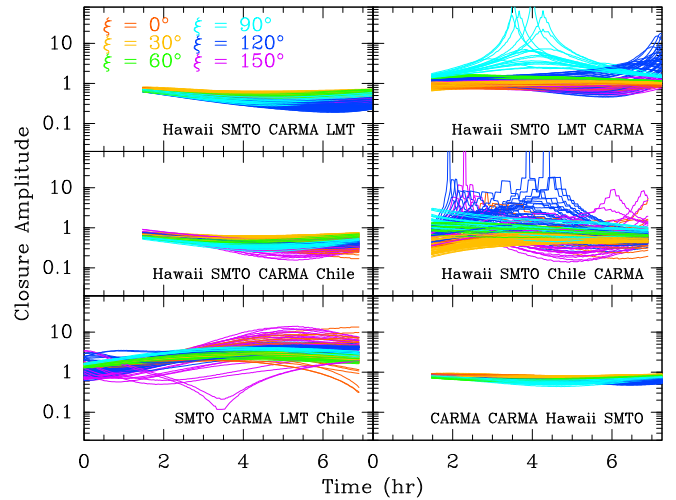}}
\caption{Predicted model closure amplitudes for all probable RIAF
  models over the time range for which Sgr~A* is above $5\degr$
  elevation at at least four of the five Western hemisphere stations.
  Colors are as in Figure~\ref{fig-baselines}.  Small jumps in the
  Hawaii-SMTO-Chile-CARMA panel are artifacts in precision when the
  Hawaii-Chile baseline passes through a null.  A single closure
  amplitude can also be formed by using two telescopes at the same
  site, such as CARMA.  Closure amplitudes are highly sensitive to
  $\xi$ and are robust against flux calibration errors.
\label{fig-closamps}
}
\end{figure}

The models predict a wide range of closure amplitudes on quadrangles
of four telescopes, as shown in Figure~\ref{fig-closamps}.  While even
a single measurement of a closure amplitude will be useful, multiple
measurements of closure amplitudes on a quadrangle with time will have
even greater power in further constraining model parameter space,
since different models can predict strongly differing closure
amplitudes over a night of observations.  As is the case for baseline
visibilities, closure amplitudes are most highly sensitive to
variations in $\xi$ among the most probable models.  From a closure
amplitude perspective, either a Chilean telescope or the LMT would
make an excellent fourth station in an observing array, although the
low predicted correlated flux densities on the Hawaii-Chile baseline
may make measurement of a second closure amplitude on an array
consisting only of Hawaii, SMTO, CARMA, and Chile difficult without
the use of phased-ALMA as the Chilean station.

An alternative observational strategy in the absence of a fourth
telescope site would be to use two different telescopes at the same
location, for instance at the CARMA site or in Hawaii.  A single
nontrivial closure amplitude can be formed from the resulting
visibilities.  The closure amplitude resulting from two CARMA
antennas, the SMTO, and Hawaii has less power to differentiate among
models than other four-antenna arrays (Fig.~\ref{fig-closamps}), but
multiple detections over a single scan may still be useful.  The
redundant information provided by the double baselines between the
CARMA site and another antenna will also aid in the detection of
fringes and reduce the systematic uncertainty in correlated flux
densities on the CARMA baselines.  Given the nondetection of Sgr~A* on
the JCMT-CARMA baseline by \citet{doeleman08}, it is likely that this
strategy would require the use of a phased-array processor at Hawaii.

\section{Conclusions}

The planned evolution of millimeter VLBI capability over the next few
years will place strong constraints on RIAF models of Sgr~A* emission.
It is important to devise an observing strategy to test the RIAF
hypothesis and extract information on the parameters of the putative
accretion disk system of Sgr~A*, including the disk orientation,
inclination, and black hole spin.  By identifying differences between
models that are consistent with present VLBI observations, we identify
four key points to guide future millimeter VLBI observations:

1.\ There is a preferred zone of baseline lengths near $3$~G$\lambda$
where there is enough angular resolution to distinguish between many
of the models but not so much resolution that a large amount of flux
is lost.  The JCMT-SMTO detections are in this zone.  The construction
of a phased-array processor for the Hawaiian millimeter telescopes
will allow for even tighter constraints to be placed on RIAF model
parameters on the Hawaii-SMTO baseline as well as increase the chance
of detecting Sgr~A* on other baselines to Hawaii.

2.\ There is a strong correlation between fluxes on the Hawaii-SMTO
and Hawaii-CARMA baselines, limiting the ability of the latter
baseline to place extra constraints on RIAF parameters unless a
phased-array processor is installed at CARMA, making Hawaii-CARMA a
more sensitive baseline than Hawaii-SMTO.  Nevertheless, the
Hawaii-CARMA measurement will be important both for ruling out simpler
models (e.g., a Gaussian or annulus) of the emission from Sgr~A* that
the present VLBI measurements cannot and for testing the RIAF
hypothesis.

3.\ VLBI arrays that include either the LMT or a Chilean telescope
provide the best constraints on RIAF models.  Predicted flux densities
on the SMTO-LMT and CARMA-LMT baselines are well in excess of 1~Jy.
Thus, the LMT can significantly enhance Sgr~A* VLBI even with a
surface not adjusted for maximum accuracy.  Likewise, since many
models predict flux densities on the SMTO-Chile and CARMA-Chile
baselines well in excess of 0.5~Jy, inclusion of even a single 10~m or
12~m Chilean telescope will result in important model constraints on
Sgr~A*.

4.\ Closure amplitudes will be critical, as amplitude calibration
errors impose systematic errors on measured baseline flux densities
that reduce their ability to constrain RIAF model parameters.  At
least four telescopes should be used in an observing array in order to
obtain closure amplitudes.  If no fourth telescope site is available
at the time of reobservation, a second antenna at the CARMA site
should be included in a $\lambda = 1.3$~mm observing array.

We note the important caveat that our analysis assumes that a RIAF
accurately and completely describes the emission from Sgr~A* at
230~GHz.  This assumption may be incorrect and is likely at best an
approximation to the actual, complex emitting region in Sgr~A*.  For
these reasons, it will be important to eventually obtain data on as
many different millimeter-wavelength VLBI baselines as is feasible.

\acknowledgments

The high-frequency VLBI program at Haystack Observatory is funded
through a grant from the National Science Foundation.

\end{document}